\documentclass[a4paper,12pt]{paper}
\usepackage[margin=2.5cm]{geometry}
\usepackage[usenames,dvipsnames]{xcolor}
\usepackage{booktabs}
\usepackage{apalike}

\usepackage{natbib}
\sectionfont{\large\sf\bfseries\color{black!70!blue}} \usepackage{rotating}
\usepackage{subcaption}
\usepackage[colorlinks,allcolors=MidnightBlue]{hyperref}
\hypersetup{
    colorlinks=true,
    linkcolor=MidnightBlue,
    filecolor=magenta,      
    urlcolor=cyan, 
    citecolor=purple,
    }
    
\usepackage[utf8]{inputenc}
\usepackage{booktabs}
\usepackage{multirow}
\usepackage{array}
\usepackage[]{graphicx} 
\usepackage{enumitem}
\usepackage[raggedright]{titlesec}

\titleformat{\paragraph}[hang]{\normalfont\normalsize\bfseries}{\theparagraph}{1em}{}
\titlespacing*{\paragraph}{0pt}{1.25ex plus 0.1ex minus .0ex}{0.05em}

\setlist{noitemsep} % or \setlist{noitemsep} to leave space around whole list

\usepackage{enumitem}
\usepackage[raggedright]{titlesec}
\usepackage{blindtext}
\usepackage{url}

\usepackage{ragged2e}
\justifying

\usepackage{authblk}

\usepackage{booktabs}
\usepackage{multirow}
\usepackage{array}
\newcolumntype{L}{>{$}l<{$}} % math-mode version of "l" column type

\RequirePackage{mathptmx} % times roman, including math (where possible)

\RequirePackage[T1]{fontenc} \RequirePackage[tt=false, type1=true]{libertine} \RequirePackage[varqu]{zi4} \RequirePackage[libertine]{newtxmath} 

\title{From Data to Actionable Understanding: A Learner-Centered Framework for Dynamic Learning Analytics}
\author{Madjid Sadallah}
\date{}

\begin{document}
\begin{center}
    \LARGE
    From Data to Actionable Understanding:\\
    
    \Large
    A Learner-Centered Framework for Dynamic Learning Analytics\\

    \vspace{1cm}
    \large Madjid Sadallah* \\  \vspace{0.5cm}
    \normalsize *Université Claude Bernard Lyon 1, CNRS, INSA Lyon, LIRIS, UMR5205, 69622, Villeurbanne, France \\ \vspace{0.5cm}

\end{center}

\begin{abstract}
Learning Analytics Dashboards (LADs) often fall short of their potential to empower learners, frequently prioritizing data visualization over the cognitive processes crucial for translating data into actionable learning strategies. This represents a significant gap in the field: while much research has focused on data collection and presentation, there is a lack of comprehensive models for how LADs can actively support learners' sensemaking and self-regulation. This paper introduces the Adaptive Understanding Framework (AUF), a novel conceptual model for learner-centered LAD design. The AUF seeks to address this limitation by integrating a multi-dimensional model of situational awareness, dynamic sensemaking strategies, adaptive mechanisms, and metacognitive support. This transforms LADs into dynamic learning partners that actively scaffold learners' sensemaking. Unlike existing frameworks that tend to treat these aspects in isolation, the AUF emphasizes their dynamic and intertwined relationships, creating a personalized and adaptive learning ecosystem that responds to individual needs and evolving understanding. The paper details the AUF's core principles, key components, and suggests a research agenda for future empirical validation. By fostering a deeper, more actionable understanding of learning data, AUF-inspired LADs have the potential to promote more effective, equitable, and engaging learning experiences.
\end{abstract}
\newpage
\section{Introduction}

The abundance of data generated in modern learning environments holds immense promise for personalized and data-driven education, yet this potential remains largely untapped. Learning Analytics Dashboards (LADs), designed to provide learners, educators, and institutions with data-driven insights, often fall short of their promise. While LADs can serve multiple stakeholders, this paper focuses specifically on their application for learners, exploring how they can be designed to empower students to translate data into actionable improvements in learning outcomes. Many current LADs prioritize data visualization over the crucial cognitive processes required for this translation \citep{Siemens2013, baker2010data}. This disconnect between data \textit{availability} and \textit{actionable understanding} can lead to frustration, underutilization of valuable learning resources, and ultimately, limit the effectiveness of data-driven educational interventions. Current LAD designs often fall short in supporting crucial interwoven cognitive processes, such as sensemaking, situational awareness, and metacognitive regulation, that are essential for translating data into action.  These processes are highly interdependent: sensemaking, the process of constructing meaning from data and experiences \citep{klein2006making, weick1995sensemaking}, relies on situational awareness—the perception and comprehension of the learning environment and its elements \citep{endsley1995toward}—and is further enhanced by metacognitive regulation—the planning, monitoring, and evaluation of one's own learning processes \citep{flavell1979metacognition}.  Existing LADs often overlook this dynamic interplay by presenting data in static formats that fail to support the iterative and contextualized nature of sensemaking. This underscores the need for a more holistic approach to LAD design.

This paper addresses this challenge by introducing the Adaptive Understanding Framework (AUF), a novel conceptual model for learner-centered LAD design. While our focus here is on learners, the principles of the AUF are broadly applicable to other LAD stakeholders. The AUF aims to bridge the gap between data availability and actionable understanding by explicitly addressing the interconnected and dynamic nature of these cognitive dimensions. Specifically, the AUF proposes an integrated approach that scaffolds users' dynamic sensemaking processes, thereby promoting actionable understanding and fostering self-regulated learning. This work is guided by the following research question: \textit{How can Learning Analytics Dashboards be designed to effectively scaffold learners' dynamic sensemaking and promote actionable understanding?}

The AUF integrates a multi-dimensional model of situational awareness, dynamic sensemaking strategies, adaptive mechanisms, and metacognitive support. This integrated approach transforms LADs from static data displays into dynamic learning partners that actively support users' sensemaking journeys. The AUF's novelty lies in its intentional and integrated design, fostering a more holistic and adaptive approach to technology-mediated learning by emphasizing the dynamic interplay between these crucial cognitive elements. This creates a personalized and adaptive learning ecosystem responsive to individual users' needs and evolving understanding.

This paper details the AUF's core principles and key components, illustrating its practical relevance with concrete examples of how its principles can inform LAD design. We analyze existing LAD frameworks, highlighting their limitations and contrasting them with the AUF's comprehensive, learner-centered approach. The paper proceeds as follows: Section 2 reviews related work on LAD design, identifying gaps addressed by the AUF. Section 3 introduces the proposed conceptual framework, detailing its core components, design principles, and theoretical underpinnings. Section 4 explores the AUF's practical application, providing guidance for designing and evaluating learner-centered LADs. Finally, Section 5 discusses the AUF's broader implications for learning and technology, including its potential to foster more equitable and engaging learning experiences, acknowledges limitations, and outlines future research directions.

\section{Related work}
This section critically examines the existing landscape of LAD design, focusing on the extent to which current frameworks support the cognitive processes essential for actionable understanding.  By analyzing prevalent approaches and identifying their limitations, we establish the need for a more holistic and learner-centered model—a gap addressed by the model presented in this paper.

\subsection{Learning analytics dashboards and cognitive perspectives}

Early LADs were primarily institution-centric, focusing on identifying at-risk students and improving retention rates. Systems like Purdue's Course Signals \citep{arnold2012course}, while valuable for prediction and risk mitigation, often prioritized data reporting over genuine learner understanding. Despite their effectiveness in identifying at-risk students, these early LADs offered limited interactivity and presented largely static data. This constrained learners' ability to engage with their own learning data and develop a deeper understanding of their progress. This highlights the need for LADs to evolve beyond mere reporting and actively engage with the cognitive processes central to learning.

Subsequent learner-centric dashboards, drawing on Self-Regulated Learning (SRL) principles \citep{winne2009supporting} and frameworks such as the ICAP framework \citep{chi2014icap}, incorporated features like goal-setting, progress monitoring, and self-reflection \citep{schwendimann2017perceiving, Jivet2021}. These systems represent a positive step towards empowering learner agency, but they are fundamentally limited by their reliance on static metrics and predefined learning pathways. Presenting progress solely based on activity completion fails to provide meaningful support for sensemaking—the active construction of understanding from complex information—restricting adaptability to learners' unique needs \citep{verbert2014learning, klein2006making}. While useful for learner agency, these approaches lack the dynamic responsiveness required for learners to adapt their strategies based on their ongoing interactions with data.

Frameworks such as the Learning Analytics Cycle \citep{Chatti2012}, offer a structured approach to data analysis, but are predominantly retrospective, lacking real-time responsiveness and dynamic adaptation necessary to effectively support learners \textit{in situ}, thus limiting their capacity to foster dynamic sensemaking and self-regulation. Similarly, Cognitive Load Theory (CLT) \citep{sweller2010cognitive}, although providing valuable insights on optimizing instructional design for individual cognition, often overlooks collaborative and adaptive dimensions essential for complex, data-driven learning. Knowledge Visualization techniques \citep{Eppler2008}, while valuable for presenting complex information, are often static and fail to dynamically support collaborative sensemaking and understanding. While individually beneficial, these approaches highlight the need for a framework capable of fostering personalized, adaptive, and real-time support for learners.

\subsection{Metacognition and self-regulated learning in learning analytics}

Effective LAD design requires a deep understanding of metacognitive and SRL dimensions, which are central to fostering learner agency \citep{zimmerman2002becoming}. SRL models \citep{Winne2009, pintrich2000role} emphasize key metacognitive processes such as planning, monitoring, and reflection. These models are instrumental in highlighting the importance of metacognition in learning, but often lack dynamic and context-sensitive mechanisms that respond to learners' evolving needs and real-time interactions with learning analytics. While SRL principles are incorporated into some dashboards, these additions are often treated as separate features rather than integral components. Consequently, existing LADs frequently lack the tools to guide real-time reflection, adaptation, and self-assessment, limiting their ability to effectively regulate learning \citep{panadero2017review}. This highlights the need for a framework that fully integrates metacognitive support within the sensemaking process.

\subsection{Collaborative learning and distributed cognition in learning analytics}

Recognizing learning as a social and collaborative process is vital for effective LAD design \citep{Hollan2000, hutchins1995cognition}. Frameworks like scenario-based approaches \citep{mohseni2024development}, while useful in collaborative learning contexts, often lack the adaptability to address the cognitive and metacognitive demands of collaborative learning in dynamic, data-rich environments. Multimodal Learning Analytics (MMLA) provides opportunities to support collaborative sensemaking by leveraging diverse data modalities \citep{Yan2024, olsen2020temporal}. MMLA has shown potential to capture collaborative interaction, but faces challenges in real-time integration, scalability, and theoretical grounding, hindering its ability to fully support collaborative and distributed cognition. Current implementations often emphasize data collection over actionable insights, which limits efficacy in fostering meaningful collaborative learning.

Distributed Situational Awareness theory \citep{Salmon2017} highlights the importance of shared cognition and the distribution of awareness across individuals and artifacts. While this theory is directly relevant, LADs rarely facilitate shared awareness or collaborative knowledge construction, often prioritizing individual over collective sensemaking. Effective LADs should adopt principles of distributed cognition to promote co-construction of knowledge and dynamic interactions among learners.

\begin{sidewaystable}
\centering
\footnotesize
\caption{Comparison of Existing Frameworks with the AUF}
\label{tab:framework_comparison}
\small
\begin{tabular}{p{3cm}p{2.5cm}p{2.5cm}p{2.5cm}p{2.5cm}p{2.5cm}p{2.5cm}p{2.5cm}}
\toprule
\textit{Framework/Model} & \textit{Dynamic Adaptability} & \textit{Personalized Sensemaking} & \textit{Integrated Metacognition} & \textit{Collaborative Sensemaking} & \textit{Distributed Cognition} & \textit{Real-time Feedback} & \textit{Multi-dimensional SA} \\
\midrule
Learning Analytics Cycle \citep{Chatti2012} & Limited (Retrospective, minimal real-time.) & Limited (No personalized sensemaking.) & Limited (No explicit metacognition.) & No (No collaboration.) & No (Individual focus.) & No (Retrospective only.) & No (No multi-dimensional SA.) \\
\midrule
SRL Models \citep{Winne2009, pintrich2000role} & Limited (General guidance, limited adaptation.) & Partial (Generic, lacks personalization.) & Partial (Metacognition, lacks real-time support.) & Limited (Individual focus, lacks collaboration.) & No (No distributed cognition.) & Emerging (Limited real-time tracking.) & No (No multi-dimensional SA.) \\
\midrule
CLT \citep{sweller2010cognitive} & No (Individual focus, no adaptation.) & No (No personalized sensemaking.) & Partial (Load reduction, no meta support.) & No (No collaboration.) & No (Ignores distributed aspects.) & No (No real-time support.) & No (No SA.) \\
\midrule
Knowledge Visualization \citep{Eppler2008} & No (Static design, no adaptation.) & No (No personalized visuals.) & Limited (No metacognitive support.) & No (Individual use only.) & No (No distributed cognition.) & No (No real-time.) & No (No SA.) \\
\midrule
Scenario-Based Frameworks \citep{mohseni2024development} & Partial (Context-based, limited beyond scenarios.) & Partial (Limited personalization, static.) & Limited (Scenario driven, not learner driven.) & Partial (Shared sensemaking, limited tools.) & Partial (Limited distributed awareness.) & Limited (Predefined feedback, no dynamic adaptation.) & No (No multi-dimensional SA.) \\
\midrule
MMLA \citep{Yan2024} & Partial (Multimodal data, limited adaptation.) & Partial (Performance insights, no personalized strategies.) & Partial (Cognitive states, no actionable support.) & Strong (Multimodal data for collaboration.) & Strong (Multimodal for distributed aspects.) & Strong (Real-time via multimodal integration.) & No (No multi-dimensional SA.) \\
\midrule
\textbf{AUF (Proposed)} & \textbf{Strong} (Adaptive to interactions, cognitive states, collaboration.) & \textbf{Strong} (Personalized paths, visuals, feedback based on profiles.) & \textbf{Strong} (Timelines, prompts, feedback, self-assessment.) & \textbf{Strong} (Shared visuals, annotations, discussions, peer review.) & \textbf{Strong} (Shared knowledge construction, collective understanding.) & \textbf{Strong} (Real-time adaptation to user and state.) & \textbf{Strong} (Clarity, Coherence, Confidence, Scope, Depth.) \\
\bottomrule
\end{tabular}
\end{sidewaystable}

\subsection{The need for a unified and dynamic approach}
Table \ref{tab:framework_comparison} highlights the strengths and limitations of existing frameworks across key dimensions such as adaptability, personalized sensemaking, and collaborative support, revealing significant gaps. These findings underscore the need for a unified, dynamic approach that integrates cognitive, metacognitive, and collaborative dimensions.

Current frameworks often address isolated aspects of learning. Multimodal Learning Analytics (MMLA) captures collaborative interactions but lacks a strong theoretical basis for interpreting data and supporting sensemaking \citep{Yan2024}. Similarly, SRL models emphasize metacognitive processes but lack real-time adaptability and collaborative support. Static frameworks like the Learning Analytics Cycle \citep{Chatti2012} and Knowledge Visualization \citep{Eppler2008} are not responsive to learners' evolving needs.

This highlights a critical gap: the absence of a holistic, adaptive framework that integrates cognitive, metacognitive, and collaborative dimensions to support real-time learner needs. A unified, dynamic framework providing personalized, real-time support is essential. The Adaptive Understanding Framework (AUF), introduced next, addresses these limitations by offering an integrated approach to support learners' evolving engagement with complex data.

\section{A conceptual model for learner-centered learning analytics}

\begin{figure}[h!]
    \centering
    \includegraphics[width=\linewidth]{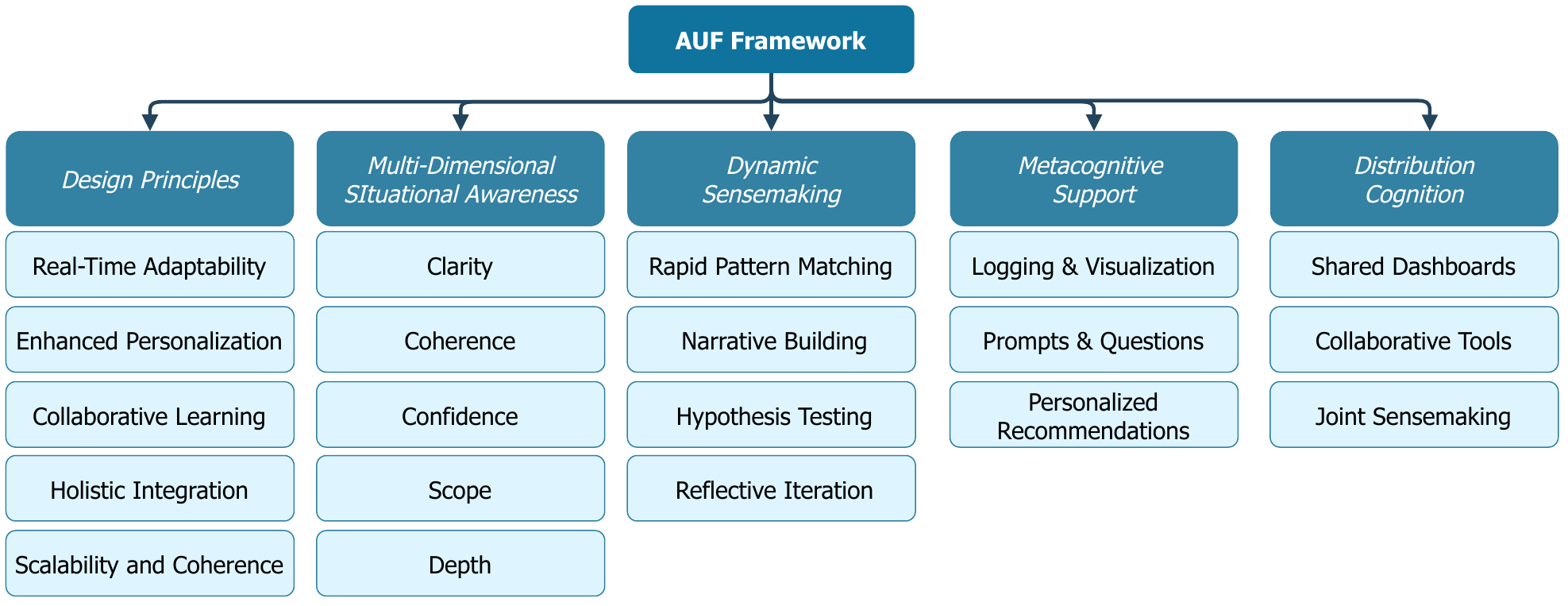}
    \caption{Overview of the Adaptive Understanding Framework.}
    \label{fig:AUF_overview}
\end{figure}

This section introduces the \textit{Adaptive Understanding Framework} (AUF), a conceptual model for learner-centered Learning Analytics Dashboard (LAD) design. The AUF addresses limitations of static data displays by integrating multi-dimensional situational awareness, dynamic sensemaking strategies, adaptive mechanisms, explicit metacognitive support, and collaborative learning. These components create a responsive learning ecosystem promoting self-regulated learning and empowering learners to make sense of their learning data. Figure \ref{fig:AUF_overview} provides an overview of the AUF.

\subsection{Core elements of the AUF: multi-dimensional situational awareness and dynamic sensemaking}

The AUF is centered on the dynamic interplay between Multi-Dimensional Situational Awareness (SA) and Dynamic Sensemaking Strategies. Unlike existing frameworks, these interconnected elements form a feedback loop: a learner's SA influences sensemaking strategy selection, and in turn, the application of these strategies shapes their SA. This iterative cycle, facilitated by the LAD, promotes deeper, more actionable understanding.

\subsubsection{Multi-Dimensional Situational Awareness}

\begin{figure}[h!]
    \centering
    \includegraphics[width=.85\linewidth]{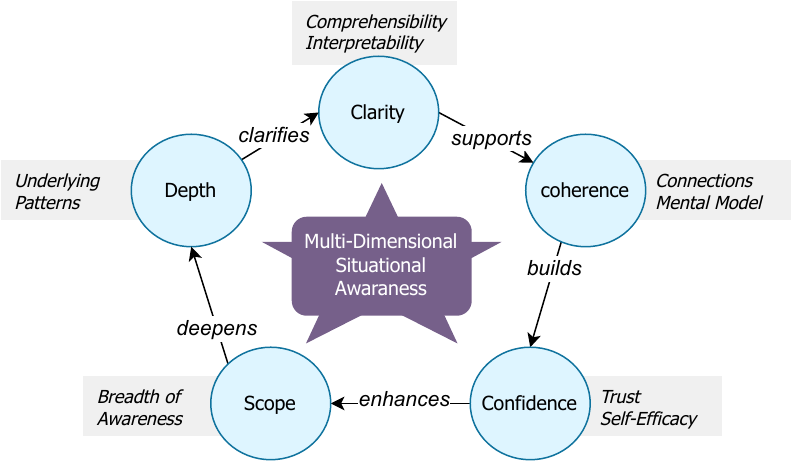}
    \caption{The AUF Multi-Dimensional Situational Awareness Model, illustrating the interconnected dimensions, and potential supporting LAD features.}
    \label{fig:multidimensional_sa}
\end{figure}

Traditional, linear models of Situational Awareness (SA) often prove inadequate for capturing the complex, dynamic, and iterative nature of learning within data-rich environments. The AUF introduces a Multi-Dimensional SA model, drawing from distributed SA theory \citep{Salmon2017}. While traditional SA models provide a foundation, they often fail to capture the collaborative, contextual, and evolving nature of learning. This theory posits that awareness is a distributed phenomenon, spanning individuals, artifacts, and the environment. As shown in Figure \ref{fig:multidimensional_sa}, the AUF's model emphasizes five interconnected dimensions – Clarity, Coherence, Confidence, Scope, and Depth – each representing a unique facet of a learner's understanding. These dimensions dynamically interact, influencing both awareness and the selection of appropriate sensemaking strategies. They also serve as critical indicators for tailored adaptive support, informing the LAD's feedback mechanisms.

\paragraph*{Clarity: ease of comprehension and reduced cognitive load}
Grounded in information processing and cognitive load theory \citep{miller1956magical, sweller2010cognitive}, Clarity, in our context, is defined by the ease with which learners perceive and interpret data presented by the LAD. High Clarity minimizes extraneous cognitive load, enabling learners to focus on meaningful data analysis rather than deciphering complex presentations. Low Clarity, in contrast, inhibits engagement and hinders effective sensemaking. LAD design for Clarity should prioritize intuitive visualizations, concise data summaries, and clear, unambiguous labels. Evaluation can be conducted using measures like response times, error rates, and cognitive load assessments, providing a direct measure of the system's efficacy in reducing cognitive strain.

\paragraph*{Coherence: integrated mental models and holistic understanding}
Rooted in sensemaking \citep{weick1995sensemaking} and constructivist learning theory \citep{bruner1960process,jonassen1999designing}, Coherence refers to the learner's ability to construct a unified mental model by integrating diverse data and concepts. This dimension is crucial for a holistic understanding of the learning experience. Low Coherence leads to fragmented perceptions, hindering pattern recognition and effective strategy development. LAD design to support Coherence should facilitate the linking of activities, concepts, and resources via tools like visual timelines and interactive annotation systems, enabling a more complete view of the learning process. Evaluation can be conducted using concept mapping and causal explanation tasks.

\paragraph*{Confidence: self-efficacy and empowered learner agency}
Drawing from self-efficacy theory \citep{bandura1986social} and metacognitive monitoring \citep{flavell1979metacognition}, Confidence, in the AUF, reflects a learner's belief in their ability to effectively navigate the LAD and the broader learning process. High Confidence promotes active engagement, whereas low Confidence leads to passivity. Crucially, the AUF aims not only to enhance confidence in abilities but also to cultivate a sense of learner agency. LAD designs should therefore represent achievements and progress, while providing self-assessment tools, and personalized resources. Evaluations can use self-reported scales, analysis of help-seeking behaviors, and tracking of system usage.

\paragraph*{Scope: awareness of broader learning context and goals}
 Drawing on Endsley's situational awareness model \citep{endsley1995toward} and situated cognition \citep{brown1989situated}, Scope describes the breadth of a learner's awareness within the learning environment. A broad Scope enables learners to connect specific learning activities to overarching goals, providing a complete understanding of the learning system. In contrast, a narrow Scope results in a fragmented and disconnected view. LAD designs should integrate visual representations of the learning environment, illustrating the connections between activities, resources, and broader career pathways. Evaluations will focus on the diversity and range of resources used, connections made across diverse learning contexts, and the range of completed learning tasks.

\paragraph*{Depth: critical data analysis and insight generation}
 Rooted in critical thinking and analytical reasoning \citep{halpern2013thought}, Depth, in the AUF, reflects the learner's ability to analyze data patterns and relationships to generate actionable insights. Low Depth leads to superficial interpretations, whereas a high Depth promotes effective strategies. This dimension is critical for actionable understanding within the learning context. LAD designs should offer tools for interactive data analysis and dynamic visualization that support data exploration and modeling, with the goal of generating meaningful learning insights. Evaluations can be conducted through tasks requiring critical data analysis, explanations of reasoning, and the complexity of the skills utilized.

The five dimensions of Situational Awareness outlined above are deeply interconnected, with progress in one often driving development in others. For instance, enhanced Clarity can lead to better Coherence, which in turn boosts Confidence, encouraging learners to explore a broader Scope and engage in more in-depth analysis. The adaptive mechanisms of the LAD are designed to support learners as they navigate these interconnected dimensions, offering personalized guidance based on their individual profiles and evolving needs. While these dimensions provide a comprehensive framework, it is important to view them as a foundation for ongoing exploration, as the complex interplay between these aspects of SA requires further investigation.

\subsubsection{Dynamic Sensemaking strategies}

\begin{figure}[h!]
\centering
\includegraphics[width=\linewidth]{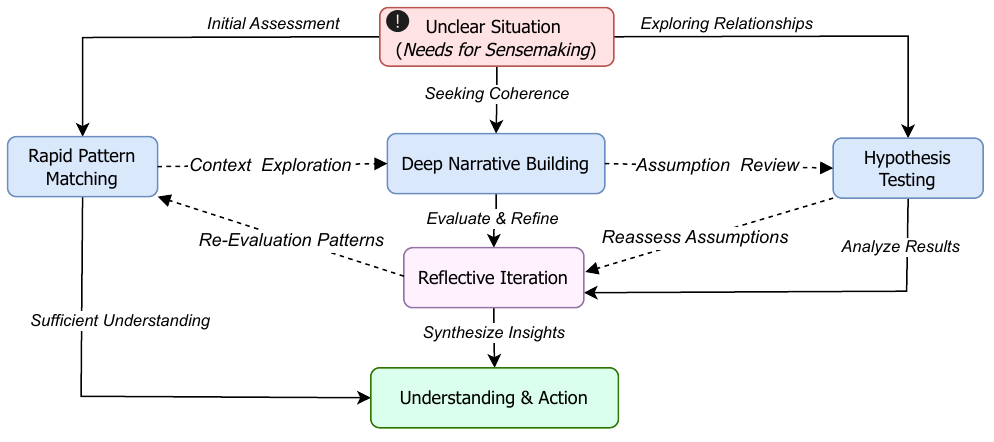}
\caption{Dynamic Sensemaking Strategies within the AUF, illustrating potential pathways and transitions between strategies as learners' understanding evolves.}
\label{fig:sensemaking_strategies}
\end{figure}

Sensemaking, the active process of constructing understanding from complex information \citep{weick1995sensemaking, klein2017sources}, is central to the AUF. The framework proposes four dynamic sensemaking strategies – Rapid Pattern Matching, Deep Narrative Building, Hypothesis Testing \& Refinement, and Reflective Iteration – representing different approaches to interpreting learning data, shaped by the learner's SA, learning goals, and engagement. While existing models often focus on specific sensemaking approaches, the AUF recognizes the importance of supporting transitions between different strategies. These strategies are not mutually exclusive or sequential, but rather flexible options that learners can transition between based on their needs. The LAD supports these transitions with feedback based on current tasks, SA, performance data, and metacognitive prompts. The AUF also recognizes learner-initiated and system-suggested transitions, promoting a personalized learning path. Reflective Iteration, aided by metacognitive support, acts as a key for continuous evaluation. Figure \ref{fig:sensemaking_strategies} illustrates these dynamic transitions.

\paragraph*{Rapid Pattern Matching: efficient identification of initial trends}
This strategy involves quickly identifying patterns in the data by activating existing mental schemas \citep{bartlett1995remembering, sweller2010cognitive}. It serves as an initial step in sensemaking, often triggered by low Clarity. While rapid pattern identification is a common first step in information processing, the AUF positions it within a broader sensemaking process, offering guidance for a more systematic approach.
Rapid pattern matching reduces cognitive load during the initial data assessment, helping learners identify key trends quickly. For example, a student might glance at a dashboard and immediately notice a consistent downward trend in their quiz scores over the past few weeks or recognize a visual pattern indicating that they tend to spend less time on assignments when they are submitted close to the deadline. The LAD supports this by offering simplified visualizations, such as line graphs or bar charts, and highlighting key trends.
Evaluating this strategy involves analyzing interactions with visuals and assessing the speed and accuracy of initial interpretations, providing insights into learners' ability to grasp the main trends in their learning data.

\paragraph*{Deep Narrative Building: personalizing learning through cohesive interpretations}
Rooted in constructivist principles \citep{jonassen1999designing}, this strategy involves creating a cohesive, personalized interpretation of learning data, linking it to prior knowledge, personal experiences, and broader goals. While current models emphasize active knowledge construction, the AUF focuses on a learner's ability to build a unique understanding by connecting new information with past experiences.
As Coherence develops, this strategy transforms raw data into a meaningful narrative that deepens understanding of the learning journey. For instance, after noticing a drop in their quiz scores, a student using Deep Narrative Building might reflect on their increased workload at their part-time job during that same period, relating this to reduced study time, and thus creating a personal narrative to explain the data. Or, a student may use the LAD to note that although their progress is good in assignments, their progress in extra learning activities is minimal, linking this with a personal preference for formal assignments and reflecting on strategies to engage with the more open activities. The LAD supports this by offering prompts, tools for annotation, and features that allow learners to build personal stories of their learning process.
Evaluating this strategy involves analyzing learner-generated explanations, concept maps, reflective logs, and assessing the quality and diversity of information incorporated into these narratives.

\paragraph*{Hypothesis Testing \& Refinement: promoting scientific reasoning and analytical skills}
Drawing on principles of scientific inquiry and the scientific method \citep{minner2010inquiry, Zimmerman2007}, this strategy involves formulating and testing hypotheses about the relationships between different aspects of learning data. It fosters critical thinking, scientific reasoning, and confidence development. While scientific reasoning is often viewed as a general skill, the AUF integrates it as a core element of sensemaking, specifically tied to interpreting learning data.
This stage activates as Depth increases, with learners becoming more confident in analyzing data. For example, after using Deep Narrative Building to relate low quiz scores to their work schedule, a student using Hypothesis Testing might formulate a hypothesis such as ``There is a correlation between the time I spend on assignments and my quiz scores.'' They would then use the LAD's tools to explore this correlation, testing this hypothesis by comparing scores with study hours. The LAD supports this by offering data visualization and statistical analysis tools, enabling hypothesis formulation and testing.
Evaluation involves observing how learners use data analysis tools, the methods they apply to test hypotheses, and the sophistication of their explanations of the relationships between data and performance.

\paragraph*{Reflective Iteration: facilitating continuous self-assessment and learning adaptation}
Grounded in metacognitive theory \citep{flavell1979metacognition, zimmerman2002becoming}, this strategy focuses on the continuous evaluation and refinement of understanding, fostering self-regulated learning and metacognitive awareness. While metacognitive theory is central to self-regulated learning models, the AUF emphasizes it as a core element of the sensemaking process, promoting ongoing self-assessment.
Reflective Iteration is crucial for long-term learning and developing a strategic approach. It allows learners to critically evaluate their learning process and refine their knowledge based on continuous feedback. For example, after testing their hypothesis about work hours and quiz scores, the student may use Reflective Iteration to realize that although work hours had an impact, their study techniques might also be contributing to lower scores. This leads to a new understanding of their learning processes, with a decision to revisit past lectures and use active learning strategies. The LAD supports this through metacognitive prompts, self-assessment tools, goal-setting mechanisms, and learning logs for reflection.
Evaluation involves examining learners' responses to metacognitive prompts, self-assessment reports, learning logs, and the frequency of requests for support.

\subsubsection{Reciprocal Interplay: the dynamic relationship between situational awareness and sensemaking}

The core of the AUF is the dynamic and reciprocal interplay between Situational Awareness (SA) and sensemaking strategies. Learners' SA directly influences their choice of strategies, while their use of these strategies, in turn, shapes and evolves their SA. This feedback loop, facilitated by the LAD, fosters a personalized learning process, where learners adapt their approaches and understanding according to their current state. Table \ref{tab:sa_sensemaking_interplay} illustrates typical links between SA dimensions and corresponding sensemaking strategies. These links reflect common patterns, though other strategies may be employed based on individual learning needs and context. The LAD supports learners by providing just-in-time guidance as their SA evolves.

\begin{table}[h!]
\footnotesize
\centering
\caption{Links between Situational Awareness and Sensemaking Strategies}
\label{tab:sa_sensemaking_interplay}
\begin{tabular}{p{1.7cm}p{1.8cm}p{4cm}p{7cm}}
\toprule
\textbf{SA state(s)} & \textbf{Most Likely Strategy} & \textbf{LAD Support} & \textbf{Example} \\
\midrule
Low \textit{Clarity} & \textit{Rapid Pattern Matching} & Simplified visualizations; quick overviews;  highlighting of key trends. & A bar chart clearly highlighting low performance in a specific assignment, with the option to expand to more details. \\
\midrule
Developing \textit{Coherence} & \textit{Deep Narrative Building} & Prompts for linking data with prior knowledge; annotation tools; visual representations of relationships. & Questions or prompts that link quiz scores with time management strategies or learning habits, encouraging learners to write down their personal interpretations. \\
\midrule
Increasing \textit{Depth} & \textit{Hypothesis Testing \& Refinement} & Statistical tools; data filtering options;  interactive visualization tools; simulation tools. & Access to a linear regression tool that will allow learners to explore causal relations between different variables, and test their own interpretations of the data.  \\
\midrule
Any SA state  &  \textit{Reflective Iteration} & Metacognitive prompts; self-assessment tools; learning logs. & Prompts that ask learners to assess their learning progress, evaluate their effectiveness, and set new learning goals.  \\
\bottomrule
\end{tabular}
\end{table}

The AUF framework's core elements, including its multi-dimensional Situational Awareness model and dynamic sensemaking strategies, establish a robust foundation for understanding the remainder of this theoretical proposal.

\subsection{Guiding design principles: operationalizing the Adaptive Understanding Framework}

This section details the five core design principles guiding the AUF and informing the design of learner-centered LADs. These principles—Real-Time Adaptability, Enhanced Personalization, Collaborative Learning, Holistic Integration, and Scalability and Coherence—operationalize the AUF elements of multi-dimensional SA and dynamic sensemaking (Section 3.1), integrating metacognitive awareness and collaborative knowledge construction to promote dynamic, personalized, and collaborative data-driven learning.

\paragraph*{Real-Time adaptability: ensuring dynamic responses to evolving learner states}
This principle emphasizes that LADs should dynamically adapt to learners' evolving SA and sensemaking strategies. While adaptive learning systems exist, they often rely on predefined pathways or static rules. The AUF, \textit{in contrast}, promotes a fluid and dynamic approach informed by learners' real-time interactions and evolving understanding. This adaptation involves adjusting visualization complexity, providing just-in-time hints, or suggesting alternative sensemaking strategies based on current needs. By responding in real time, the LAD better supports dynamic learning paths, aligning with learners' shifting Clarity, Coherence, Confidence, Scope, and Depth, and chosen sensemaking strategies. Real-time adaptation is facilitated through rule-based systems and machine learning \citep{sutton2018reinforcement} that collect and analyze data in real time, enabling dynamic adjustments.

\paragraph*{Enhanced personalization: tailoring the learning experience to individual profiles} 
This principle champions tailoring LADs to individual differences in SA and sensemaking for a more relevant and engaging experience. While existing systems strive for personalization, they are often limited by generic learner profiles and predefined learning paths. The AUF, however, proposes adaptive learner profiles that capture individual learning preferences and characteristics dynamically, moving beyond static representations. This includes recommending appropriate sensemaking approaches based on the learner's profile and current context, and providing tailored feedback. This principle guides the LAD to adapt to individual learning paths. This personalized approach aims to optimize diverse learning needs, using machine learning for recommendations \citep{linden2003amazon, ricci2010introduction} and Natural Language Generation (NLG) \citep{gatt2018survey} for tailored feedback.

\paragraph*{Collaborative learning: fostering shared sensemaking and distributed cognition}
Recognizing the social nature of sensemaking and the importance of shared SA, this principle promotes shared data exploration, annotation, and discussion within the LAD. Although collaborative learning has been a focus of research in learning technologies, LADs rarely incorporate features that promote true collaborative knowledge construction. The AUF extends current LADs by integrating tools that facilitate shared data exploration and co-construction of knowledge, going beyond simple data sharing. The aim is to create opportunities for learners to engage in collaborative inquiry, promoting diverse perspectives and enriching individual sensemaking processes. By enabling collective data exploration, meaning co-construction, and insight sharing, the LAD becomes a hub for distributed cognition, facilitating both individual and shared SA development. This can be achieved through synchronous and asynchronous communication tools, shared interactive dashboards, and peer review/co-creation features, fostering a collaborative learning community.

\paragraph*{Holistic integration: promoting cohesion and seamless functionality within the lad}
This principle underscores the synergy between all AUF components, stressing the need for cognitive, metacognitive, and collaborative elements to work seamlessly, supporting learners' evolving SA and sensemaking strategies. Current educational technologies often focus on separate cognitive, metacognitive, and collaborative dimensions, treating them as isolated aspects of learning. The AUF, \textit{in contrast}, seeks to dynamically integrate these elements into a cohesive learning experience. This creates a unified and adaptive learning environment that responds dynamically to both individual and shared sensemaking. Integration can be achieved through a unified architecture and seamless data integration across all LAD modules, using learner data to inform every aspect of the system and ensuring a cohesive and interconnected user experience.

\paragraph*{Scalability and coherence: designing for broad applicability}
This principle advocates for accessible personalized support across diverse educational contexts. While flexibility is a recognized need in educational technologies, many systems lack the coherence to provide personalized support at scale. The AUF addresses this by proposing a modular architecture and a consistent design, allowing for broader implementation without compromising the user experience. By prioritizing modularity and consistent design, the framework adapts to various learning environments while maintaining a coherent experience. The goal is to apply the AUF's principles flexibly, supporting diverse learning styles and contexts, catering to individual and collective SA development, and varied sensemaking approaches. This can be realized through modular design, cloud-based infrastructure, and standardized APIs, facilitating seamless deployment and accommodating diverse needs and constraints.

\begin{table}[h!]
    \centering
    \footnotesize
    \caption{Interconnectedness of the AUF Design Principles}
    \label{tab:principles_interconnectedness}
    \begin{tabular}{p{2.2cm}p{2.2cm}p{2.2cm}p{2cm}p{2cm}p{2cm}}
        \toprule
         & \textit{Real-Time Adaptability} & \textit{Enhanced Personalization} & \textit{Collaborative Learning} & \textit{Holistic Integration} & \textit{Scalability \& Coherence} \\
        \midrule
        \textit{Real-Time Adaptability} &  & {Responsive Feedback} & {Collaborative Adaptations} & {Data-Driven Integration} & {Flexible Design} \\
        \midrule
        \textit{Enhanced Personalization} & {Adaptive Paths} & & {Personalized Interaction} & {Tailored Experiences} & {Equitable Learning} \\
        \midrule
        \textit{Collaborative Learning} & {Shared Dynamic Data} & {Collaborative Insights} & & {Community Integration} & {Context Flexibility} \\
       \midrule
       \textit{Holistic Integration} & {Integrated Responsiveness} & {Integrated Personalization} & {Integrated Collaboration} & & {System Consistency} \\
        \midrule
        \textit{Scalability \& Coherence} & {Adaptable Framework} & {Scalable Personalization} & {Scalable Collaboration} & {Unified Architecture} & \\
        \bottomrule
    \end{tabular}
\end{table}

\paragraph*{Connectedness of the AUF principles} 
The interconnectedness of these principles, illustrated in Table \ref{tab:principles_interconnectedness}, creates a robust and adaptable framework for effective learning experiences. The table emphasizes that the principles are not isolated aspects, but rather mutually reinforcing elements that create a cohesive system. For instance, the intersection of \textit{Real-Time Adaptability} and \textit{Enhanced Personalization} indicates that the system's adaptation mechanisms must be tailored to the learner's individual profile and their particular needs, leading to a more effective feedback mechanism. Similarly, the intersection between \textit{Collaborative Learning} and \textit{Holistic Integration} highlights the importance of creating a learning environment in which collaborative tools are dynamically integrated with all aspects of the learning process, allowing for a dynamic and more cohesive shared understanding. This interconnectedness highlights the AUF's holistic and dynamic nature, emphasizing the importance of considering all these principles in the design and implementation of technology-mediated learning environments.

\subsection{Theoretical foundations: grounding the AUF in established learning principles}

This section examines the theoretical underpinnings of the AUF, highlighting its unique contributions by articulating its basis in established learning theories. It also explores how the framework leverages distributed cognition to inform collaborative learning experiences, contributing to a more conceptually robust approach for technology-mediated education.

\subsubsection{Integrating cognitive, metacognitive, and collaborative perspectives}

The AUF offers key theoretical contributions to learner-centered LAD design. {While existing frameworks draw on cognitive, metacognitive, or collaborative principles individually}, the AUF proposes an integrated model of dynamic Situational Awareness (SA) and sensemaking. This emphasizes the reciprocal relationship between a learner's understanding of their context and their active interpretation of information. {This unified perspective goes beyond current approaches}, providing a theoretical basis for adaptive support systems that respond to learners' evolving understanding. It also outlines a conceptual approach for personalized support that accommodates diverse learning styles and sensemaking strategies. By proposing integrated metacognitive scaffolding, the framework empowers self-regulated learning by embedding mechanisms for active reflection and deeper understanding, going beyond separate prompts or add-ons. Additionally, it frames collaborative sensemaking as a key driver of SA, recognizing the social nature of learning and the importance of shared understanding. The AUF also bridges the gap between complex data and actionable understanding, empowering learners to make informed decisions, refine strategies, and improve learning. These contributions highlight the innovative and theoretically grounded nature of the framework. {These integrated theoretical aspects define the AUF as a holistic framework merging diverse educational perspectives}.

\subsubsection{Distributed Cognition: conceptualizing the LAD as a hub within a learning ecosystem}

\begin{figure}[h!]
\centering
\includegraphics[width=\linewidth]{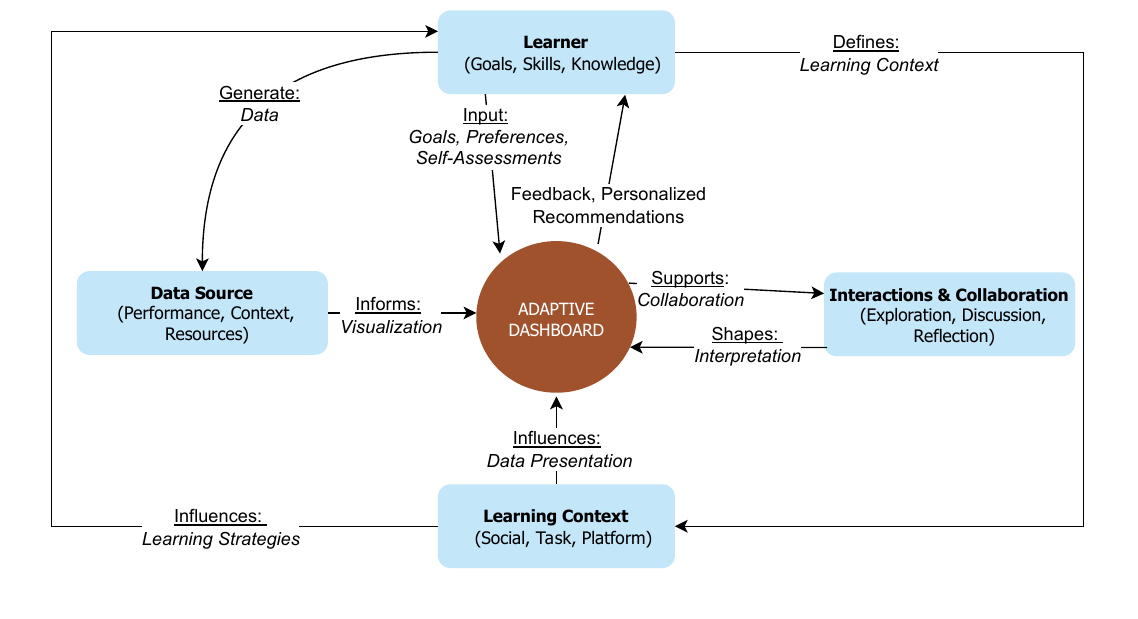}
\caption{The LAD as a Conceptual Hub in a Distributed Cognitive System.}
\label{fig:interactions}
\end{figure}

The AUF conceptualizes learning as inherently distributed across learners, data, tools, and socio-cultural contexts. {While distributed cognition theory has gained traction}, its application to LAD design has been limited. Drawing upon this theory \citep{Hollan2000, hutchins1995cognition}, the AUF envisions the LAD as a conceptual hub within a distributed cognitive ecosystem (Figure \ref{fig:interactions}), challenging the traditional focus on individual cognition and shifting attention towards collaborative knowledge construction. Here, the LAD acts as a mechanism facilitating interactions and promoting shared sensemaking.

Within this ecosystem, knowledge construction is a shared process, encompassing individual and collaborative activities. Learners engage in individual sensemaking using personalized LAD features for reflection, and in collaborative sensemaking using the LAD for shared inquiry, contributing to the collective knowledge of the learning community. This dual focus creates a personalized and interconnected learning environment, acknowledging both individual agency and collective interaction, and recognizing the distributed nature of learning. The LAD acts as a mediator, facilitating a dynamic interplay between individual and collective understanding.

\paragraph*{Facilitating shared knowledge construction in a distributed cognitive system}
The AUF framework promotes distributed cognition by highlighting core features. The system promotes transparency by externalizing thought processes, using mechanisms like shared visualizations with annotations to support collective understanding. Moreover, the system promotes perspective-taking by enabling learners to share and compare interpretations of data, creating platforms for structured dialogues. Additionally, it facilitates knowledge negotiation using shared tools for collaborative hypothesis testing, supporting evidence-based reasoning. Finally, collaborative activities promote collective SA through peer feedback and co-created narratives, supporting knowledge co-construction and shared understandings of the learning context.

\paragraph*{Illustrative example: collaborative data analysis in a smart city design project}

Consider, as a conceptual example, urban planning students using an AUF-inspired LAD to analyze smart city transportation data. Some students might focus on exploring traffic patterns, initiating discussions and sharing interpretations using Rapid Pattern Matching. Others might engage in constructing narratives through Deep Narrative Building by connecting bottlenecks to different transit options. Another group of learners could explore improved bus routes using Hypothesis Testing. Shared tools would allow learners to co-construct knowledge, while individual strategies provide a pathway for personalized learning. Throughout this process, a metacognitive layer would offer personalized feedback and prompts, while the LAD's Scalability and Coherence would enable effective collaboration. {This example serves to illustrate how the framework promotes learning in a collaborative, dynamic, and data-driven setting, going beyond what is offered by current, individualized educational tools}.

\subsubsection{Adaptive mechanisms}
The AUF emphasizes dynamic adaptation to individual learners' needs and evolving understanding through core mechanisms that translate cognitive principles into interactive LAD features. Unlike existing frameworks, which often highlight adaptivity, the AUF provides a conceptual approach grounded in cognition and real-time learner-system interactions. Guided by Real-Time Adaptability and Enhanced Personalization, it proposes mechanisms to dynamically respond to learners' evolving SA, sensemaking strategies, metacognitive reflections, and collaborative interactions, offering conceptual possibilities for future implementation.

\paragraph*{Adaptive data presentation: customizing visualizations to support evolving understanding}
The AUF proposes data presentation as a dynamic process that adapts to the learner's cognitive state. {While many visualization techniques exist}, they often lack the flexibility to adapt to diverse learner characteristics. The AUF envisions adaptive visualization, linking data presentation to a learner's evolving SA. Visualizations are tailored to a learner's SA and sensemaking strategy, adjusting types and detail levels to guide attention and enhance understanding. For example, a novice learner with low Clarity would be presented with simplified visualizations, transitioning to more complex representations as their Scope and Depth increase.

\paragraph*{Guided narratives: contextualizing feedback with interactive learning stories}
The AUF also explores interactive, guided narratives for contextualized support. {While guided narratives have been shown to be useful in many learning contexts}, their implementation in data-driven learning often lacks adaptive mechanisms. Narratives are tailored to a learner's SA and sensemaking strategy, offering personalized guidance and scaffolding. For example, a learner with low Confidence might be presented with a simplified timeline, gradually revealing more details as their Confidence increases. The framework may also offer alternative learning resources aligning with their current learning style.

\paragraph*{Feedback loops: ensuring dynamic responses to learner actions and sensemaking}
The AUF integrates feedback loops as dynamic mechanisms directly linked to learners' actions, sensemaking strategies, and evolving understanding. {While feedback is a core part of learning processes}, its implementation is often static, not responsive to real-time needs. The system offers tailored feedback addressing misunderstandings and guiding learners toward more appropriate strategies. This feedback can incorporate adaptive guidance, linking learners to relevant resources or personalized explanations.

\paragraph*{Dynamic level of detail: balancing cognitive load with progressive information disclosure}
To prevent cognitive overload, the AUF proposes a dynamic level of detail, adjusted based on a learner's SA. Although Cognitive Load Theory highlights managing cognitive load, the AUF offers a conceptual approach for a dynamic application, beyond general guidelines. Information complexity begins with simplified representations, revealing more details as Clarity and Confidence increase. The framework also promotes transitions to complex sensemaking strategies, prompting learners to form and test their interpretations of data.

\paragraph*{Personalized recommendations: providing contextually relevant guidance and support}
The AUF envisions personalized recommendations based on learner profiles, learning context, and real-time behavior. While recommender systems are common in technology-mediated education, their capacity to adapt based on real-time information about learner interactions is often limited. For instance, a learner may receive recommendations that align with their current understanding, or alternative learning paths suited to their style. Recommendations are presented with clear explanations to empower learners to take control of their learning.

\paragraph*{Explainable AI (XAI): fostering transparency and building learner agency}
To enhance transparency and promote learner agency, the AUF proposes Explainable AI (XAI) techniques. Although XAI is becoming more common, its application in educational technologies is still lacking. XAI mechanisms clarify the reasoning behind adaptive decisions, enabling learners to understand how the system responds to their needs. For example, the system might explain which data points or interactions triggered a specific recommendation, supporting metacognitive awareness, transparency, and trust.

\section{Operationalizing and evaluating the AUF: a framework for design and assessment}

This section articulates how the Adaptive Understanding Framework (AUF) can serve as a guide for the design of learner-centered LADs. It further presents a structured approach for evaluating LADs through the AUF lens and proposes a research agenda to explore the impact of AUF-inspired designs. This section aims to establish the AUF as a foundational framework for informed design, rigorous assessment, and impactful research within technology-mediated learning.

\subsection{Using the AUF to guide learner-centered LAD design}

The AUF is a conceptual framework that provides a methodology for designing LADs that empower learners as active, informed, and strategic users of their learning data. Its core principles—multi-dimensional situational awareness (SA), dynamic sensemaking, adaptive support, metacognition, and collaboration—are interconnected elements that inform design decisions. The following sections illustrate how the AUF can guide the development of effective learner-centered LADs.

\subsubsection{Structuring data and interactions for meaningful learning}
he AUF provides a comprehensive framework for structuring both data and user interactions within LADs to foster meaningful learning experiences.  This approach emphasizes the dynamic interplay between data presentation and learners' active engagement in sensemaking.

The AUF's multi-dimensional situational awareness model offers a detailed perspective for organizing learning data, enhancing learners' understanding of their progress, learning context, and the overall learning environment.  Guided by the principle of \textit{Clarity}, LADs should prioritize clear, concise, and relevant information presentation, minimizing cognitive load through intuitive visual encodings, consistent labeling, effective use of whitespace, and progressive disclosure.  To foster \textit{Coherence}, LADs should facilitate the construction of integrated mental models of learning through interactive timelines, concept maps, and narrative construction tools that connect disparate learning elements.  Promoting learner \textit{Confidence} requires clear, actionable feedback highlighting progress, alongside self-assessment tools and access to support resources.  Expanding the \textit{Scope} of learners' perspectives involves connecting specific learning activities to broader learning goals, curricular structures, and potential career pathways through system overviews and integrated resource hubs.  Finally, fostering \textit{Depth} in learning necessitates LAD features that support in-depth data analysis and interpretation, enabling learners to filter, manipulate, and explore data from diverse perspectives, thereby fostering critical reasoning and analytical skills.

Beyond data structuring, the AUF's dynamic sensemaking framework emphasizes that LADs should actively support learner engagement with complex information.  Adaptive scaffolding is crucial for guiding learners' transitions between sensemaking strategies, dynamically adjusting visualization complexity and offering tailored support based on individual learner profiles and progress.  Furthermore, integrating interactive tools that support annotation, concept mapping, hypothesis formulation, and narrative construction enables active interpretation and meaningful knowledge construction, dynamically adapting to learner needs and fostering deeper engagement with the learning process.

\subsubsection{Active knowledge construction through dynamic interactions}

The AUF posits that LADs should not merely present data passively, but should actively facilitate learners' construction of knowledge through dynamic and responsive interactions. This perspective emphasizes that sensemaking is not a static process, but rather an active, iterative, and context-dependent endeavor. To this end, the AUF proposes specific design principles to support active knowledge construction:

Firstly, LADs should incorporate \textit{adaptive scaffolding} to facilitate seamless transitions between diverse sensemaking strategies. This entails dynamically adjusting the complexity of data visualizations, offering tailored prompts and feedback, and providing contextually relevant recommendations for effective data interpretation based on a learner's individual profile, their current progress, and their demonstrated strategies. This responsiveness is critical for supporting learners through the complexities of engaging with rich learning data.

Secondly, the integration of \textit{interactive tools} is paramount to enhance the meaning-making process. These tools should move beyond simple visualization techniques, enabling learners to actively engage with data through annotation, concept mapping, hypothesis formulation, evidence-based argumentation, and narrative construction. Furthermore, these tools should not be static but should dynamically adapt to the learner’s actions and needs, actively supporting the iterative and evolving nature of sensemaking. The provision of such tools is vital for fostering active interpretation, deep understanding, and the personalized construction of meaningful knowledge from learning data. This also requires the LAD to support mechanisms of reflection to allow learners to integrate the newly constructed knowledge into their learning strategies.

Finally, and crucially, LADs bridge data exploration with actionable insights. They must not only support the exploration of data, but they must also promote the integration of data-derived conclusions into learning practices. This might include prompts for reflective practice, tools that link interpretations to learning goals, or features that encourage the development of revised learning strategies based on data-driven insights. By making explicit the link between exploration and action, the AUF framework aims to empower learners to become active agents in their own learning. This requires not just tools to explore data, but tools that support the process of transforming information into knowledge and new learning strategies.

\subsubsection{Design considerations for effective and responsible LADs}

The AUF underscores the critical role of personalization, metacognition, collaboration, and ethical data practices in designing effective and responsible learning environments.  These considerations are not supplementary features but are integral aspects of a holistic approach to data-driven learning.

The AUF posits that LAD designs should implement \textit{adaptive mechanisms} to personalize the learning experience based on individual learner profiles, learning goals, and engagement patterns. This requires tailoring information presentation and offering responsive feedback, grounded in real-time data analysis, integrating personalization with metacognitive processes and diverse learning strategies to better fit the needs of each individual learner.

Further, drawing from the AUF's emphasis on metacognition, LADs should promote \textit{self-regulated learning} by providing learners with tools and features that enhance their metacognitive awareness. This includes the seamless integration of reflective prompts, self-assessment tools, and personalized feedback mechanisms that support learners in monitoring their learning, making strategic choices, and becoming self-directed learners. The integration of metacognitive awareness should be organic, a part of the core experience.

Acknowledging the importance of social learning, the AUF framework calls for LAD designs to include features that support \textit{collaborative learning} by enabling shared data exploration, collaborative annotation, and dynamic discussions, actively fostering shared sensemaking and knowledge construction through meaningful interaction.

Finally, the AUF emphasizes \textit{ethical data handling} as a core design consideration. LADs must prioritize transparency in data collection, ensure informed consent, and implement robust privacy and security measures. This is crucial to protecting learner data, mitigating algorithmic biases, and promoting fairness and equity within the learning environment, safeguarding both the data and the learner.

\subsection{Evaluating LADs through the AUF lens}

The AUF provides a structured approach for evaluating both existing and AUF-inspired LADs. This evaluation approach moves beyond basic usability assessments, focusing instead on how well a system supports learners' cognitive, metacognitive, and social processes essential for actionable understanding.  The following core questions, derived from the AUF's key tenets, should guide this evaluation process:

\begin{itemize}
    \item \textit{Situational Awareness:} How effectively does the LAD enhance learners' situational awareness across the dimensions of clarity, coherence, confidence, scope, and depth?
    \item \textit{Sensemaking:} To what extent does the LAD support learners in effectively applying diverse sensemaking strategies, enabling personalized and meaningful interpretations of learning data?
    \item \textit{Adaptive Personalization:} How do the LAD's adaptive mechanisms contribute to a personalized learning experience, empowering learners to control their learning paths and self-regulate their interactions with data?
    \item \textit{Metacognition:} How effectively does the LAD foster metacognitive awareness, supporting self-regulated learning and informed strategic decision-making?
    \item \textit{Collaboration:} Does the LAD facilitate collaborative learning and knowledge co-construction, promoting shared understanding and diverse perspectives?
    \item \textit{Ethical Considerations:} How does the LAD address ethical considerations regarding data collection, usage, privacy, and algorithmic bias, promoting responsible and equitable data-driven learning?
\end{itemize}

These evaluative questions can be addressed through a range of robust methodological approaches, including heuristic evaluations leveraging the AUF as a comprehensive framework for systematically assessing LAD designs; user testing and think-aloud protocols to gain direct insights into learner experiences and cognitive processes; mixed methods research combining quantitative data (e.g., performance metrics, log data, validated instrument scores) with qualitative data (e.g., interviews, observations) to holistically evaluate LAD impact; and comparative studies employing rigorous experimental and quasi-experimental designs to contrast AUF-inspired LADs with traditional approaches and evaluate design decisions' impact on learning metrics.  The selected methods must be carefully aligned with the specific research questions and target audience, acknowledging and mitigating any inherent methodological limitations to ensure a balanced and rigorous evaluation.  The AUF, therefore, offers a powerful lens for scrutinizing LAD efficacy and advancing the design of learning analytics systems that genuinely support meaningful, data-driven learning.

\subsection{A research agenda for investigating the impact of AUF-inspired designs}

The proposed framework provides a theoretically grounded foundation for future research into data-driven learning. This research program is designed to empirically investigate the real-world implications of the AUF, specifically examining how design choices influence learner engagement, the learning process, and learning outcomes. The overarching goal is to validate the practical utility of the AUF and contribute to a deeper understanding of effective technology-mediated learning environments.

To guide this empirical work, our research will be structured around several core research questions exploring key facets of the AUF's design and impact:

\begin{itemize}
    \item How do design choices guided by the AUF's multi-dimensional model of situational awareness (encompassing clarity, coherence, confidence, scope, and depth) influence learners' interaction with data and their capacity for self-regulated learning?
    \item How do LAD features designed to support the dynamic sensemaking strategies, as defined by the AUF, affect learners' ability to construct meaning from complex learning data and subsequently impact learning outcomes?
    \item How do adaptive mechanisms, informed by the AUF, influence learners' self-regulated learning behaviors, their sense of ownership over their learning paths, and their perceived agency within the learning environment?
    \item How do metacognitive prompts and support tools, implemented in alignment with the AUF, affect learners' metacognitive awareness, their strategic approach to learning, and their capacity to adapt to diverse learning contexts?
    \item To what extent do collaborative learning tools, inspired by the AUF, support knowledge co-construction, meaningful interaction, and shape the overall learning experience?
    \item How do design choices, informed by ethical data handling practices, influence learners' trust in the system, their engagement with data-driven learning, their perceptions of fairness, and the equitable access to learning opportunities?
\end{itemize}

To address these interconnected research questions, a mixed-methods approach, integrating multiple research methodologies, will be employed.  Design-Based Research (DBR) \citep{brown1992design} will be utilized to iteratively explore the AUF's practical application through the design, implementation, and evaluation of LADs within authentic learning environments. Experimental and quasi-experimental studies, incorporating control groups, will investigate causal relationships between specific design features informed by the AUF and demonstrable learning outcomes.  These quantitative methods will be complemented by qualitative data collection and analysis, including interviews, think-aloud protocols, and observations, to provide a richer, more contextualized understanding of how AUF-inspired designs impact learners.  This mixed-methods approach will provide a comprehensive perspective on the complexities of implementing the AUF in real-world educational settings.

Data collection will employ a range of validated instruments and techniques.  Psychometrically validated instruments will assess self-efficacy, motivation, metacognition, and learners' overall approach to learning. Performance-based assessments will evaluate learners' analytical skills, interpretive abilities, and critical thinking.  Log data analysis will provide insights into learner interaction patterns, feature utilization, and data exploration strategies.  Furthermore, qualitative data analysis will elucidate learners' perceptions of the system, perceived challenges, and overall user experience.  Triangulating these diverse data sources will provide a robust and nuanced evaluation of the effectiveness and impact of AUF-inspired LAD designs.

\section{Discussion and conclusion}
This section discusses the AUF's core contributions, its limitations, and future research directions. We emphasize the framework's potential to advance the field of learning analytics while acknowledging the practical and ethical considerations essential for responsible implementation.

\subsection{Core contributions}
The AUF offers a novel approach to LAD design, shifting the focus from data visualization to the dynamic interplay of cognitive, metacognitive, and collaborative processes. While existing approaches have contributed valuable insights by focusing on specific dimensions, such as static presentations of data \citep{Sedrakyan2020} or individualized learning pathways \citep{arnold2012course}, the AUF prioritizes their holistic integration, supporting learners in the active construction of understanding from complex data \citep{weick1995sensemaking, klein2006making}. The core contributions include the following.

\paragraph*{A learner-centered approach to active sensemaking} 
The AUF promotes \textit{active sensemaking}, transforming LADs from static data displays into interactive tools that facilitate learners' personalized engagement with data. This contrasts with many current systems that, while providing valuable data summaries \citep{Liu2024}, may not adequately support the active cognitive work necessary for deep understanding.

\paragraph*{Holistic personalization through adaptive mechanisms} 
The AUF proposes a nuanced approach to personalization, encompassing adaptive mechanisms that dynamically respond to learners' evolving situational awareness and sensemaking strategies. This creates a more individualized experience, adjusting to needs, preferences, and learning styles in real-time. This nuanced approach to personalization moves beyond static profiles and predefined learning paths common in existing systems \citep{linden2003amazon, ricci2010introduction}, recognizing the dynamic and iterative nature of learning \citep{Winne2009}.

\paragraph*{Metacognitive scaffolding for enhanced self-regulation} 
The AUF emphasizes the importance of metacognitive scaffolding that is integrated into the core learning process, not as an added feature. This approach promotes self-regulated learning by supporting learners in planning, monitoring, and reflecting on their learning. While self-regulated learning is acknowledged as a crucial element in educational models \citep{zimmerman2002becoming}, the AUF attempts to strengthen the link between self-regulation and data-driven learning by embedding metacognition within the core sensemaking loop.

\paragraph*{Collaborative knowledge construction through a social perspective} 
The AUF positions collaborative sensemaking as a key driver of understanding. It promotes knowledge co-construction within a distributed perspective, emphasizing social interaction as essential for both individual and shared SA development \citep{hutchins1995cognition, Hollan2000}. This approach acknowledges a significant gap in current LADs, which often lack features that facilitate shared understanding and distributed cognition \citep{Salmon2017}, and that may limit the possibilities for shared learning and reflection.

\paragraph*{Data-informed decision-making for all stakeholders} 
While centered on learners, the AUF's emphasis on transparency and adaptability supports data-informed decision-making for all stakeholders, including educators and institutions, fostering more collaborative and equitable approaches to technology-mediated learning.  This moves beyond learner-centric applications, promoting a more participatory and democratic use of learning analytics.

\subsection{Significance and broader implications}
The AUF offers a theoretically grounded and learner-centered approach to LAD design, with broader implications for learning analytics and technology-mediated learning.  By emphasizing the dynamic interplay of cognitive and social processes, drawing upon cognitive science, distributed cognition theory, and adaptive learning principles, the AUF provides a framework for empowering all stakeholders in the educational process.  It has the potential to promote data literacy and equitable access to learning, potentially mitigating the ``digital divide'' \citep{van2014digital} and fostering a more democratic and participatory approach to learning analytics.

Specifically, the AUF can facilitate data-informed decision-making for learners, educators, and institutions. For learners, it fosters improved self-regulated learning through integrated metacognitive scaffolding, encouraging active engagement with data to promote deeper understanding, increased motivation, and stronger data literacy skills.  This active engagement can lead to improved learning outcomes and a greater sense of ownership over the learning process.  For educators, the AUF enables data-informed instructional design, personalized interventions, and the promotion of collaborative learning, providing them with actionable insights to enhance pedagogical practices.  Institutions can leverage the AUF to monitor program effectiveness, identify areas for improvement, and foster a data-driven culture of continuous improvement.

Furthermore, the AUF's principles extend beyond standalone LADs, offering a flexible and adaptable model for enhancing a range of educational technologies, including AI-powered tutoring systems, collaborative learning platforms, and learning management systems.  By incorporating the AUF's focus on sensemaking, situational awareness, and metacognitive regulation, these technologies can be designed to better support learners' cognitive needs and promote deeper, more meaningful learning experiences.  This adaptability positions the AUF as a valuable contribution to the ongoing evolution of technology-mediated learning.

\subsection{Limitations and future directions}
While the AUF offers a promising approach to learner-centered LAD design, it is essential to acknowledge its limitations and outline future research directions.

\subsubsection{Practical and technical challenges for implementation}

Implementing AUF-inspired LADs may encounter practical and technical challenges. These include the need for robust data interoperability across different learning platforms, the computational demands of real-time adaptation and personalized feedback, and the complexity of designing accessible user interfaces that cater to a wide range of learner abilities and technological expertise. Furthermore, integrating multiple data sources seamlessly, while ensuring data privacy and security, adds a layer of complexity that requires careful consideration \citep{boyd2014critical}. These issues present significant but not insurmountable challenges, that need to be carefully considered for practical implementation. While existing frameworks have often focused on specific aspects of design or data handling, such as basic reporting or data visualization \citep{arnold2012course,Verbert2014}, the AUF will need to explore how to solve these complex issues.

\subsubsection{Ethical and social considerations for responsible use}

The ethical implications of implementing the AUF require ongoing consideration to ensure the framework's power is used responsibly and equitably. These include the importance of ensuring transparency and user control over data collection, safeguarding learners' motivation and self-determination, prioritizing data privacy and security \citep{nguyen2024lifelong}, and proactively addressing potential biases in underlying algorithms. Unfair and discriminatory outcomes, derived from biased data and algorithms, may hinder the equitable possibilities of the framework if not addressed proactively \citep{kizilcec2022algorithmic}. While these ethical considerations are commonly acknowledged, their practical application within a dynamic and adaptive system, such as the AUF, will need further exploration, especially in terms of transparency, learner empowerment, and the equitable design of complex learning environments. Future work should also explore how the AUF's principles can be integrated with humanistic design approaches, that can help to address some of the possible social implications of the framework.

\subsubsection{Future research directions}
To comprehensively explore the potential of the proposed framework, rigorously refine its theoretical underpinnings, systematically address its inherent limitations, empirically investigate its real-world impact, and iteratively develop its core principles through evidence-based scholarship, our future research will be guided by several interconnected lines of inquiry.

A foundational prerequisite is the development and implementation of empirically grounded LADs explicitly designed according to the theoretical tenets of the AUF. This initial phase will necessitate meticulous attention to the dynamic interplay of situational awareness, sensemaking, and metacognitive regulation, providing the tools for future investigation.

With functional AUF-inspired LADs established, the subsequent priority will be creating psychometrically sound and validated assessment instruments. These instruments must assess the impact of designed LADs on learners' situational awareness, sensemaking proficiency, and metacognitive capabilities, enabling comparisons of the efficacy of AUF-inspired LADs against extant systems. These evaluations will focus on the specific cognitive processes modulated by the AUF, differentiating it from approaches emphasizing data visualization \citep{Sedrakyan2020} or self-regulated learning without real-time scaffolding \citep{Winne2009}.

To gain a nuanced understanding of the framework's impact, empirical investigations, employing rigorous quantitative and qualitative methodologies, will be conducted across diverse subjects and learner populations. These investigations will detail the LADs' influence on learning outcomes, self-regulated learning competencies, and learner engagement, with longitudinal studies evaluating long-term effects on learning behaviors and motivation. This work will also examine the framework's adaptability to diverse learning profiles.

Recognizing the importance of cultural context, future research will address cultural adaptation and sensitivity, ensuring resulting AUF-inspired LADs are accessible and equitable across diverse contexts. This entails a commitment to culturally responsive design principles that acknowledge the heterogeneity of learning styles.

In tandem with these explorations, investigations will explore the synergistic potential of integrating AUF-inspired LADs with emergent educational technologies, notably AI-powered tutoring and advanced multimodal environments \citep{Yan2024}. These explorations will prioritize optimizing these technologies for personalized learning and data-driven student engagement, while maintaining the learner's role and promoting human-driven sensemaking.

To establish the framework's empirical contributions, comparative analyses will benchmark AUF-inspired LADs against established frameworks and educational tools, employing validated techniques to elucidate the strengths and weaknesses of the AUF when contrasted with alternatives, such as scenario-based design frameworks \citep{mohseni2024development}, regarding learning outcomes, user experience, and scalability.

Finally, rigorous evaluation of potential cognitive overload from the complexity of AUF-inspired LADs will be prioritized. This evaluation will focus on data presentation and interface design to ensure that the system's benefits outweigh potential information fatigue, informing design strategies balancing complex systems with comprehensible and actionable interfaces guided by human-centered design.

\subsection{Conclusion}
The Adaptive Understanding Framework provides a theoretically robust roadmap for the design and evaluation of data-driven Learning Analytics Dashboards. By prioritizing learners' cognitive and social needs, and by fostering metacognitive awareness and self-regulated learning competencies, the AUF aims to advance more equitable and effective approaches to technology-mediated education.  The framework synthesizes insights from cognitive science \citep{miller1956magical}, distributed cognition \citep{hutchins1995cognition,Hollan2000}, and collaborative learning \citep{dillenbourg1999you}, forging an interdisciplinary foundation for future exploration and innovation within the field of learning analytics.

Translating the AUF's conceptual architecture into practical application necessitates a commitment to rigorous ongoing evaluation, iterative adaptation, and adherence to ethical principles that place the learner at the center of the design process. The AUF, as a flexible and adaptable framework, is intended to invite collaborative engagement from researchers, designers, and educators, supporting the development of more inclusive, meaningful, and potentially transformative learning experiences. This collaborative approach holds the potential to empower learners and contribute to a more equitable and humanistic future for technology-enhanced education.

\small
\bibliographystyle{apalike}
\bibliography{references}
\end{document}